\documentclass{article}

\usepackage{xcolor}
\definecolor{mygreen}{RGB}{0, 176, 80}
\definecolor{myblue}{RGB}{0, 112, 192}
\definecolor{mypurple}{RGB}{112, 48, 160}

\usepackage{arxiv}

\usepackage[utf8]{inputenc} 
\usepackage[T1]{fontenc}    
\usepackage{hyperref}       
\usepackage{url}            
\usepackage{booktabs}       
\usepackage{amsfonts}       
\usepackage{nicefrac}       
\usepackage{microtype}      
\usepackage{lipsum}
\usepackage{graphicx}
\usepackage{authblk}
\usepackage{textcomp, gensymb}
\usepackage{float}
\graphicspath{ {./images/} }
\usepackage{mathtools}
\usepackage{amsmath,amssymb}
\usepackage{comment}
\usepackage{mathrsfs}
\usepackage{graphicx,nicefrac}
\usepackage{float}
\usepackage{amssymb}

\usepackage{comment}
\usepackage{scalerel}
\usepackage[overload]{empheq}
\usepackage{textgreek}

\usepackage{algorithm}
\usepackage{algpseudocode}
\usepackage{enumitem}

\usepackage{mathrsfs}
\usepackage{graphicx,nicefrac}

\usepackage[utf8]{inputenc}
\usepackage[english]{babel}

\usepackage[margin=1cm]{caption}
\usepackage{comment}
\usepackage{scalerel}
\usepackage[overload]{empheq}

\usepackage{float}

\usepackage{mathtools}

\usepackage{authblk}
\title{Imaging across multiple spatial scales with the multi-camera array microscope}

\author[1,+]{Mark Harfouche}
\author[2,+]{Kanghyun Kim}
\author[2]{Kevin C. Zhou}
\author[2]{Pavan Chandra Konda}
\author[1]{Sunanda Sharma}
\author[3]{Eric E. Thomson}
\author[4]{Colin Cooke}
\author[2]{Shiqi Xu}
\author[2]{Lucas Kreiss}
\author[2]{Amey Chaware}
\author[2]{Xi Yang}
\author[2]{Xing Yao}
\author[2]{Vinayak Pathak}
\author[3]{Martin Bohlen}
\author[1]{Ron Appel}
\author[1]{Aurélien Bègue}
\author[1]{Clare Cook}
\author[1]{Jed Doman}
\author[1]{John Efromson}
\author[1]{Gregor Horstmeyer}
\author[1]{Jaehee Park}
\author[1]{Paul Reamey}
\author[1]{Veton Saliu}
\author[3]{Eva Naumann}
\author[1,2,4,*]{Roarke Horstmeyer}

\affil[1]{Ramona Optics Inc., 1000 W Main St., Durham, NC 27701, USA}
\affil[2]{Department of Biomedical Engineering, Duke University, Durham, NC 27708, USA}
\affil[3]{Duke School of Medicine Department of Neurobiology, Durham, NC 27708, USA}
\affil[4]{Department of Electrical and Computer Engineering, Duke University, Durham, NC 27708, USA}

\affil[*]{Corresponding author: roarke.w.horstmeyer@duke.edu}
\affil[+]{These authors contributed equally to this work}

\begin{document}
\maketitle
\begin{abstract}
This article experimentally examines different configurations of a novel multi-camera array microscope (MCAM) imaging technology. The MCAM is based upon a densely packed array of ``micro-cameras'' to jointly image across a large field-of-view at high resolution. Each micro-camera within the array images a unique area of a sample of interest, and then all acquired data with 54 micro-cameras are digitally combined into composite frames, whose total pixel counts significantly exceed the pixel counts of standard microscope systems. We present results from three unique MCAM configurations for different use cases. First, we demonstrate a configuration that simultaneously images and estimates the 3D object depth across a $100$ x $135$ $mm^2$ field-of-view (FOV) at approximately 20 $\mu$m resolution, which results in 0.15 gigapixels (GP) per snapshot. Second, we demonstrate an MCAM configuration that records video across a continuous $83$ x $123$ $mm^2$ FOV with two-fold increased resolution (0.48 GP per frame). Finally, we report a third high-resolution configuration (2 $\mu$m resolution) that can rapidly produce 9.8 GP composites of large histopathology specimens.
\end{abstract}

\section{Introduction}
A general challenge in the design of optical microscopes is to identify strategies that overcome a fundamental trade-off between imaging resolution and field-of-view (FOV). Optical microscopes range from FOVs of several centimeters when resolving at multi-\textmu{m} resolution, to less than a millimeter when imaging at sub-\textmu{m} resolution~\cite{zheng2014fourier}. The total number of spatial points resolvable by a standard optical microscope, commonly referred to as the imaging system space-bandwidth product (SBP)~\cite{lohmann1996space}, is generally between 10 and 50 million (10-50 megapixels)~\cite{park2021review}.

There are a number of compelling reasons why new microscope techniques with increased SBP can be more useful than conventional microscopes. For example, a large-SBP microscope would enable observation of multiple model organisms, such as zebrafish (\textit{D. rerio}), fruit flies (\textit{D. melanogaster}), and nematodes (\textit{C. elegans}) during natural movement over a large area~\cite{kim2017pan, nguyen2016whole, johnson2020probabilistic, krishnamurthy2020scale}. In addition, a high SBP system would assist with the rapid inspection of large electronics components~\cite{acciani2006application, baek2004inspection}, semiconductor wafers~\cite{wu2014wafer}, microfluidic systems~\cite{castiaux2019review} and various materials~\cite{chen2020solar, yun2020automated} during manufacturing. Finally, high SBP microscopes would also facilitate novel parallelized assays to increase the throughput of high-content imaging and screening experiments common to the fields of pharmacology, toxicology and drug discovery~\cite{Boutros2015}.

\begin{figure}[t]
    \centering
    \includegraphics[scale = 0.32]{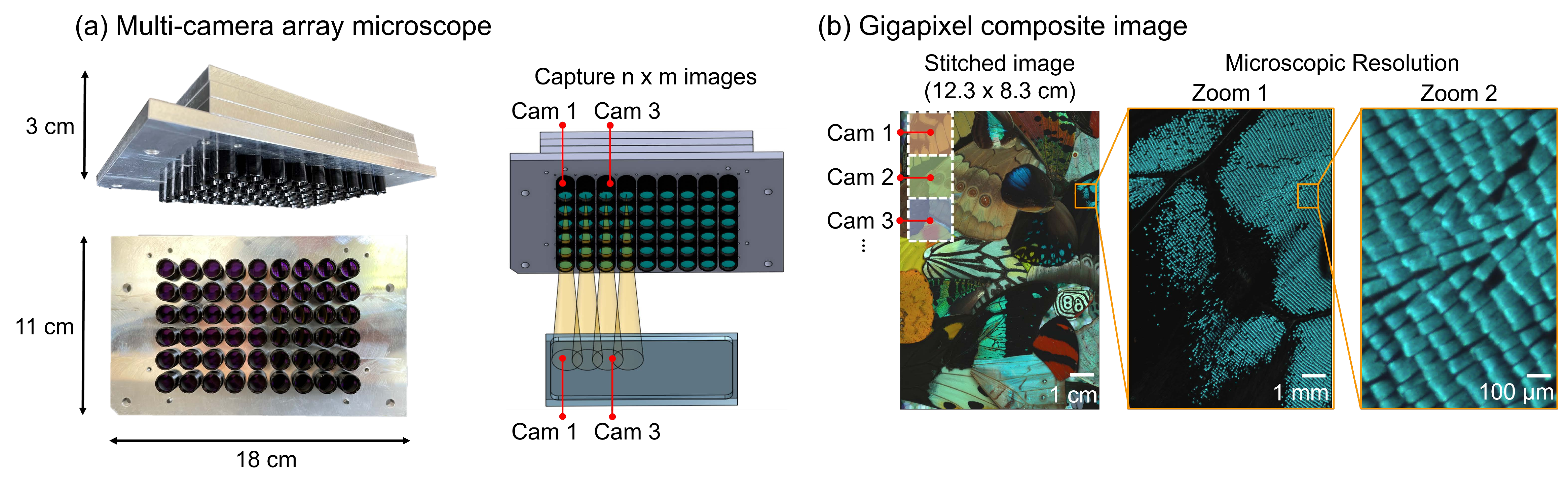}
    \caption{The multi-camera array microscope (MCAM) for large field-of-view, high-resolution imaging at video rate. (a) Photo of MCAM imaging unit containing a 6 x 9 array of individual 13 megapixel CMOS sensors and associated lenses (0.7 gigapixel per snapshot), with diagram of operation. (b) Example image, created as a stitched composite of 54 simultaneously acquired images from unique, slightly overlapping sample areas.}
    \label{fig:teaser}
\end{figure}

Unfortunately, large-SBP imaging cannot be easily achieved by directly increasing the FOV of a microscope objective lens capable of a desired imaging resolution. As specified by the lens scaling law~\cite{lohmann1989scaling}, all lenses are affected by optical aberrations, and the size of optical aberrations scale linearly with the diameter of a lens. Large-diameter lenses must utilize additional optical elements to correct for aberrations to maintain a fixed resolution over a desired FOV. Additional elements must be added in a super-linear manner~\cite{lohmann1989scaling,Brady:18parallel}, which leads to a rapid increase in size, weight, complexity and cost of large-SBP microscope objective lenses~\cite{schniete2018fast}. At the same time, the largest currently available image sensors only contain several hundred megapixels~\cite{canon_global}, which presents a second challenge to directly scaling up standard single-lens microscope designs.

Instead of relying upon a single extremely large objective lens, most current microscopes utilize mechanical scanning to overcome typical FOV limits. Scanning systems capture multiple high-resolution images in a step-and-repeat manner, which are then tiled together into a final large-area composite. Scanning microscopes are available from a number of companies and form the foundation of modern wide-area inspection and whole-slide imaging (WSI)~\cite{farahani2015whole}. Unfortunately, mechanical scanning is inherently slow. For example, a single 96 well plate (8 x 12 cm² in size) typically requires approximately 8 minutes to scan~\cite{chan201996, maioli2016time}. In addition, while sequential scanning works well for static samples, it is not an option for rapidly moving samples.

An alternative to mechanical scanning is to scan a microscope's illumination while capturing images in a time-sequential manner. Structured illumination microscopy~\cite{gustafsson2000surpassing} and Fourier ptychography~\cite{zheng2013wide,Konda:20FPReview} process such variably illuminated images into final, high-resolution image composites. Similar strategies have also been employed utilizing speckle illumination~\cite{mudry2012structured}. Time-sequential measurement of stochastic or photoactivated emission is also standard practice in super-resolution microscopy~\cite{rust2006sub,manley2008high}. Finally, alternative methods have utilized scanning spots produced by microlens arrays~\cite{Orth:12,Pang:12}, as well as steering the illumination across large specimens via a parabolic mirror~\cite{ashraf2021random}, to observe a larger sample area. While such approaches offer various merits over step-and-repeat scanning systems, the above methods still must capture multiple images over time, and thus cannot easily record large area, high-resolution synchronized video of dynamic objects.

A multi-aperture design can circumvent the lens scaling law by utilizing many lenses and digital sensors to record images in parallel~\cite{brady2012multiscale}. The first demonstrations of digital multi-aperture imaging originated in standard cameras designed for imaging distant objects and can be generally summarized in two categories. The first ``direct'' category uses multiple cameras arranged in a compact array with no additional collection optics. This design was explored to decrease optical form-factor~\cite{tanida2001thin}, to implement light-field capture~\cite{wilburn2005high}, to adopt hierarchical modular designs~\cite{Yuan:21}, and is now a common feature in many smartphone camera systems. The second category of approach adopts a cascaded or "multi-scale" lens design in which an array of cameras is focused upon an intermediate image, typically produced by a large, primary lens ~\cite{Brady:09, Tremblay:12, brady2012multiscale, marks2014characterization}. This latter cascaded strategy has been implemented in several interesting configurations~\cite{Schuster:19}, including a microscope that produces images at sub-micrometer resolution over approximately one square centimeter~\cite{fan2019video}. However, there has been limited work to date examining the direct use of multiple imaging systems to record image and video across a continuous area at high spatial resolution in parallel for applications in microscopy, which is the focus of this work. 

There are a variety of benefits of direct arrays for microscopic imaging applications. For example, small optics exhibit fewer aberrations, and thus simpler, more compact, and less expensive lenses can be used within the array design. Similarly, smaller and inexpensive CMOS pixel arrays are now made in large quantities for the smartphone camera market. Put together, these two insights point to a relatively simple and inexpensive wide-area microscope design. In this work, we demonstrate the ability to use such an array to directly record images at micrometer-scale resolution over a FOV of several hundred square centimeters in parallel. Our primary design, which we term a multi-camera array microscope (MCAM), includes 54 micro-cameras that are integrated into a regularly packed grid with a 13.5-mm center-to-center spacing. In the following, we cover several configurations of this new technology to achieve different functionalities: the ability to record 3D snapshots across a 100 x 135 mm² area at approximately 20$\mu$m full-pitch lateral resolution and 42$\mu$m axial sensitivity, video across a continuous 83 x 123 mm² area at 2X higher lateral resolution, and high-resolution image with 9.8 GP over a similar area. As detailed below, each of these configurations is a function of selected imaging magnification that one can easily adjust using the same hardware.

\section{Multi-camera array microscope design} 
\label{Multi-camera array microscope design}
\begin{figure}[!h]
    \centering
    \includegraphics[scale = 0.44]{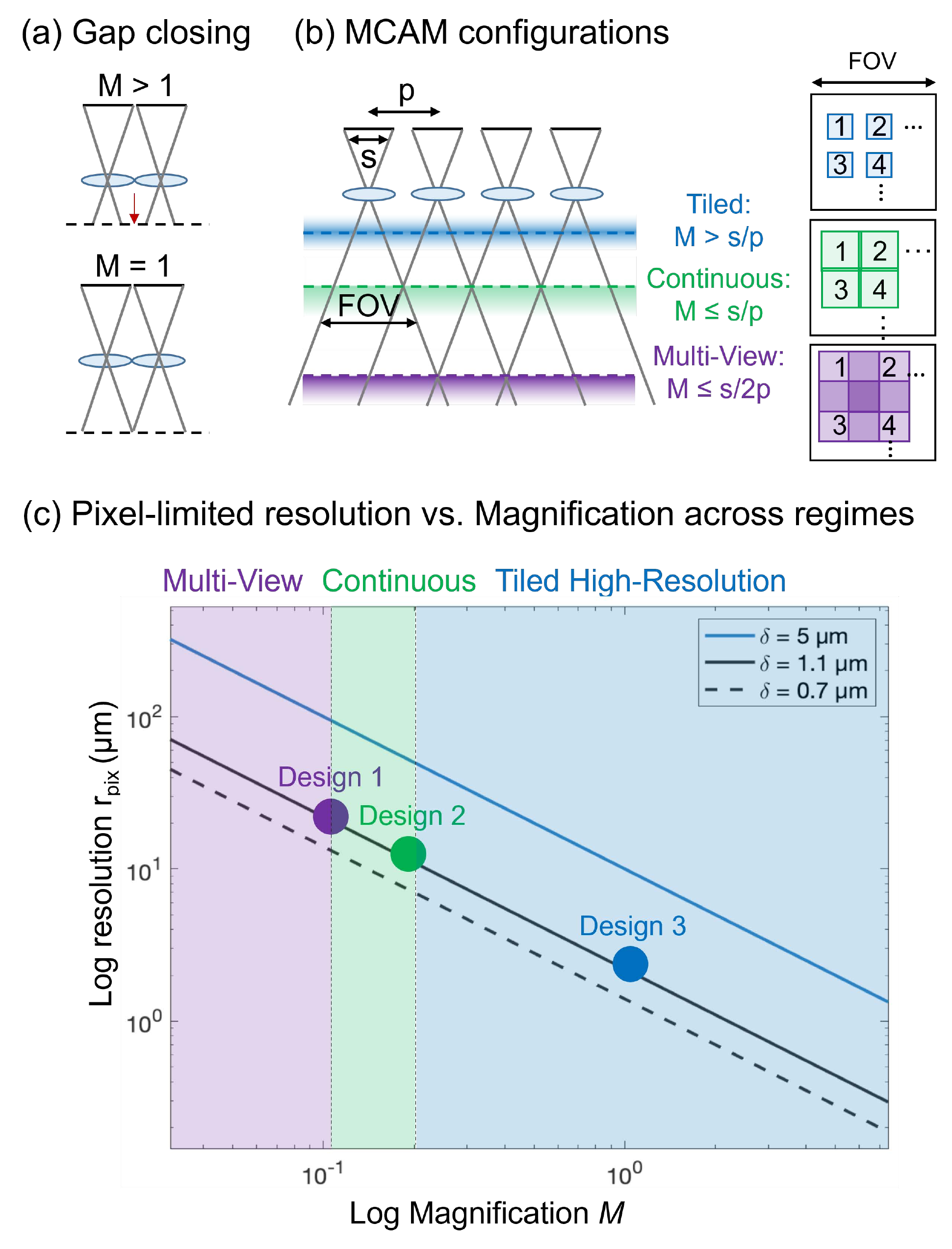}
    \caption{(a) Diagram of two co-planar microscopes with immediately adjacent sensors and lenses. A magnification $M>1$ leads to an unobserved gap (red arrow), which is closed when $M=1$. (b) 1D MCAM diagram with sensor pitch $p$ larger than sensor width $s$. The MCAM can be configured to operate in three regimes (Tiled, Continuous, Multi-View) depending upon selected magnification. (c) MCAM pixel-limited resolution as a function of magnification for several common pixel sizes (log-scale). Regimes of operation are noted at the top and experimental demonstrations are marked with colored circles.}
    \label{fig:threemodes}
\end{figure}

We will begin by deriving several key properties, such as the relationship between magnification and pixel-limited resolution, for direct-array imaging systems. Our starting assumption is that the array of interest is planar, contains identical imaging systems that are arranged in a uniform grid, and is imaging a large two-dimensional specimen of interest. We will scrutinize these assumptions in later sections. As we aim to configure the entire array, and thus each lens-sensor pair, to directly image at micrometer-scale resolution, we will loosely refer to each lens-sensor pair as a microscope. It is useful to begin by considering how only two such individual microscopes side-by-side can be configured to image a continuous area (i.e., such that the FOV of each microscope abuts the other, see Figure~\ref{fig:threemodes}(a)). Naturally, one would aim to place the microscopes as close as possible to one another. At the extreme limit, the sensors for each microscope would ideally sit immediately adjacent to one another.

Assuming this limit is achievable, we need to consider the objective lens placed in front of each sensor. As shown in Figure~\ref{fig:threemodes}(a), it is clear that a lens with a magnification $M > 1$ will result in a gap between FOVs of adjacent microscopes. While it is possible to uniquely tilt each lens and sensor to minimize or remove this gap~\cite{konda2018parallelized}, such a configuration introduces several sources of experimental complexity that hinder large array development. Instead, one can simply use a lens with $M \leq 1$ to ensure that the FOVs of each imaging system directly touch one another (Figure~\ref{fig:threemodes}(a)). This leads to a first key property of direct array microscopy with a planar array - the maximum magnification must be less than unity to ensure continuous imaging across an extended FOV.

At first glance, this key property would appear problematic for achieving high-resolution imaging. In such low-magnification conditions, the finite sensor pixel size is critical for the overall resolution. The system's pixel-limited full-pitch resolution $r_{pix}$ is easily found by projecting the finite pixel width $\delta$ onto the sample plane as $r_{pix}=2\delta/M$. Historic CCD and CMOS sensors contained pixels on the order of $\delta = 5-10 \mu$m in width, which suggests that even in an ideally tiled array, the minimum full-pitch resolution would be $10-20 \mu$m at best when configured for continuous coverage, which precludes many applications. Over the past years, however, the average pixel size of CMOS sensors has decreased dramatically. It is now common to find CMOS pixels in the range of $\delta = 0.7 - 1 \mu$m~\cite{kim20205}. Alternative sensor designs likewise include even smaller pixel widths~\cite{fossum2016quanta}. Modern CMOS pixel arrays, typically found in most smartphone cameras, achieve a minimum full-pitch resolution $r_{pix}=2\delta/M$ between 1.4 and 2 $\mu$m when using a magnification of M = 1. This approximately matches the resolution offered by a current standard 4X or 10X objective lens with numerical apertures in the 0.15-0.3 range.

In practice, placing individual image sensors immediately adjacent to one another is challenging (in part due to mechanical packing and electrical routing constraints). Considering an image sensor of an active area width $s$ in one dimension, this implies that the inter-camera pitch $p$ must satisfy $p > s$ in practical MCAM implementation (Figure~\ref{fig:threemodes}(b)). The sensor pitch $p$ also \emph{defines the lateral separation} between the optical axis of the adjacent systems in a planar array design. This defines the physical separation between the center of each image FOV. To fulfill our aim of imaging a contiguous surface without any gaps, we must ensure that for each lens-sensor pair in the array, the imaged FOV is equal to the sensor pitch: $FOV=p$ (see Figure~\ref{fig:threemodes}(b), green dashed line). When configured to image across a continuous object plane without any gaps, the magnification of each imaging system, defined as the ratio between its image FOV and an object FOV, is then given by:
\begin{equation}
M=\frac{s}{p}.
 \label{magnification_continous}
\end{equation}
Inserting this required magnification into our definition of the imaging system's pixel-limited full-pitch resolution now provides a relationship in terms of sensor width $s$ and pitch $p$:
\begin{equation}
r_{pix}=\frac{2\delta}{M}=\frac{2\delta p}{s},
 \label{pixel_res}
\end{equation}
for a configuration to image a continuous surface. Of course, the imaging system magnification can also be selected to be greater than or less than Eq.~\ref{magnification_continous}, which will impact the total FOV coverage of the array microscope and its resolution. We will detail these alternative configurations below. We plot pixel-limited resolution as a function of magnification and for several common sensor pixel widths in Figure~\ref{fig:threemodes}(c), which also highlights the three configurations of the MCAM investigated here.

The resolution of each imaging system within the array is also impacted by its optics. Based on the specific pixel-limited resolution set by Eq.~\ref{pixel_res}, it is useful to match the optical resolution of each single lens to this limit. In practice, this condition is relatively easy to satisfy for the compact imaging optics used by each lens-sensor pair, as we demonstrate for a number of configurations and as theoretically examined in prior work~\cite{marks2011close}. Each of the micro-camera imaging lenses is relatively small, since the lens diameter must be less than $p$, which is often on the order of 1 cm. Furthermore, small lenses ensure tight array packing. Following the lens scaling law outlined in Ref.~\cite{lohmann1989scaling}, such small-diameter lenses are easily designed to offer high performance (i.e., minimal aberrations) across a wide range of specifications. While there are certainly optical limits on achievable resolutions and FOVs for each micro-camera, a primary benefit of utilizing a multi-aperture imaging arrangement is the ability to avoid many of the challenges associated with large lens design. In other words, since the required SBP of each micro-camera remains relatively small, designing a lens to meet pixel-limited resolution requirements is much easier (and cheaper) than attempting to create a large lens to capture a much larger SBP.

Finally, the number of individual lens-sensor pairs within the entire array needs to be considered. This number is simply a function of the desired total FOV for the array system. In this work, we first demonstrate a primary configuration that covers an approximate 100 x 135 mm² area using a 6 x 9 array of 54 micro-cameras. We also demonstrate the ability to tile together four individual multi-camera arrays to directly extend this total FOV by a factor of four. This straight-forward scalability of the total FOV is a unique benefit of arrayed microscopy, such as the MCAM system (see also Figure~\ref{fig:extension_multipleboards}).

The selected magnification of all lens-sensor pairs within the array is a key design choice that drives three primary regimes of operation (see Figure~\ref{fig:threemodes}(b)). We summarize these three unique operation regimes next, before experimentally demonstrating their benefits and trade-offs.

\subsection{Configuration 1 - Multi-View Imaging}
Using a large working distance and a low image magnification, the MCAM can be configured for "multi-view" imaging. In this configuration, the FOVs of individual lens-sensor pairs overlap so that each location in the sample is imaged by at least two unique micro-cameras (see Figure~\ref{fig:threemodes}(b), purple). As we will show, a multi-view imaging geometry offers a number of interesting benefits. For instance, this allows stereoscopic imaging to estimate the depth of imaged objects, or to apply photogrammetry software to jointly reconstruct a height map of the imaged object while stitching together the final large-area composite. Alternatively, spectral or polarimetric multiplexing may be utilized to capture additional specimen information from such overlapped imagery. In the extreme scenario of all $N$ micro-cameras within the array imaging the same object location, which is only possible when the object distance is increased to a large stand-off distance, the micro-camera array captures a light-field-type dataset ~\cite{levoy2006light}, from which depth can be estimated~\cite{broxton2013wave, prevedel2014simultaneous, lin2015camera, pei2019occluded}. We have found that a useful MCAM configuration is at the opposite limit, where a small amount of inter-camera overlap between immediately adjacent sensors can produce high-quality depth estimation, while the entire array can still yield a significantly increased imaging FOV.

The magnification of an MCAM configured for multi-view imaging typically satisfies,
\begin{equation}
M_m \le \frac{s}{2p} \le \frac{1}{2}.
 \label{magnification_design1}
\end{equation}
which indicates that each micro-camera FOV has a width of $2p$ to ensure that one point on the sample plane maps to at least two micro-cameras (except at the boundaries). This will ensure that in two dimensions every point is imaged by 4 micro-cameras with square image sensors. For example, inserting the condition $M_m = 1/4$ ($p = 2s$) and a pixel size of $\delta=1.1 \mu$m into $r_{pix}=2\delta/M$ yields an 8.8 $\mu$m lower bound for full-pitch resolution. This approximately matches the resolution provided by a 1.25X microscope objective lens (with 21 mm FOV diameter) commonly used for macroscopic (2D) inspection~\cite{zheng2014fourier}. The number of micro-cameras required to image a desired total sample plane FOV area $A$ in a multi-view configuration is $\lceil$$A/4p^2$$\rceil$. We note that parts of the FOV of the corner micro-cameras may not include more than one viewpoint per sample plane area, which must be taken into account during post-processing analysis.

\subsection{Configuration 2 - Continuous FOV Imaging}
The Continuous FOV regime is entered when the magnification is increased past the multi-view scenario, but the entire surface of an object is still viewed by at least one micro-camera. This configuration has higher spatial resolution than the multi-view geometry and requires the following magnification:
\begin{equation}
M_c \le \frac{s}{p} \le 1.
 \label{magnification_design2}
\end{equation}
The continuous FOV configuration was the primary focus of our explanation at the beginning of Section~\ref{Multi-camera array microscope design}, where we explained how a continuous area (i.e., without any gaps in the FOV) is observed by a planar array when its magnification is less than one. Typically, a small amount of inter-camera image overlap (approx. 5-10\%) is required for effective composite stitching. 

\subsection{Configuration 3 - Tiled High-Resolution Imaging}
To obtain even higher image resolution, one can increase the magnification of each single micro-camera such that a gap appears between adjacent FOVs (Figure~\ref{fig:threemodes}(a)):
\begin{equation}
M_t > \frac{s}{p}.
 \label{magnification_design3}
\end{equation}
Such a "tiled" configuration no longer images across a continuous sample area in one snapshot. Instead, it captures data from a discrete, non-contiguous sample area with a FOV width of $s/M$ along one dimension. The achievable resolution of such a configuration depends on the selected magnification and eventually the imaging optics. Based on $r_{pix}=2\delta/M$ and assuming $M_t = 2$ and a $\delta = 1.1$ $\mu$m pixel size, we can estimate the full-pitch pixel-limited resolution to 1.1 $\mu$m for a tiled imaging arrangement, which matches the resolution achieved by a standard 10X microscope objective lens.

In a tiled imaging configuration, the MCAM operates similarly to a large number of individual microscopes configured to image in parallel~\cite{wilburn2005high}. To observe a macroscopic area, mechanical scanning of either the specimen or the array can be used to fill in the FOV gaps over multiple image acquisitions. The number of scan locations required to fill in the FOV gaps across the entire specimen plane in one dimension is $\lceil$$pM/s$$\rceil$ (i.e., the number of micro-camera FOVs of size $s/M$ that fit within an inter-camera spacing $p$). Assuming square image sensors and a square micro-camera packing geometry, $($$\lceil$$pM/s$$\rceil$$)^2$ unique scan locations must be visited during imaging.


Compared to standard step-and-repeat imaging with a single-lens microscope, an array with $N$ micro-cameras can image a specified surface at least $N$ times faster, since it has to scan $N$ times fewer locations, to increase imaging throughput accordingly. A second benefit is a greatly reduced scanning travel range. A tiled imaging MCAM only has to mechanically scan over the inter-camera spacing distance $p$ to fill in missing FOV gaps, as opposed to standard microscopes which must scan across the entire extent of the desired aggregate FOV. In other words, $1/N$ less movement is required. In applications where the throughput is important, this can result in rather dramatic savings, e.g., in imaging of large histology slides or when inspecting large semiconductor wafers for defects.

\section{Results}
To test each of the three imaging configurations, we constructed a prototype MCAM system containing a 6 x 9 array of micro-cameras. The 54 individual 13 megapixel CMOS sensors (ONSemi AR1335, 3120 x 4208 pixels, $\delta =1.1~\mu$m pixel width) were tiled at a $p=13.5$ mm pitch on a single PCB board. We designed an optomechanical mount to hold a 6 x 9 array of customized lenses (25.05 mm effective focal length, f/4, 13 mm outer diameter, fabricated by Edmund Optics). These lenses were separated by the same $p=13.5$ mm pitch. Each lens can be focused individually via a custom thread mount. To test all three imaging configurations, we used the same MCAM design with all lenses focused to a common plane, and adjusted the distance between the lenses and sensors by moving the optomechanical mount of the lens array. For tiled imaging (configuration 3), an additional 6 x 9 array of lenses was included to minimize aberrations (see details below). The sample working distance was controlled via a 3-axis stage.

Image data from all micro-cameras was routed to a single Field-Programmable Gate Array (FPGA) before transmission to a single desktop computer via a PCIe link. The FPGA allows control over the settings of all 54 camera sensors, e.g., exposure time and gain, through a global address. Once configured, a single command initiates synchronized image or video acquisition across all image sensors (approx. 0.7 GP per snapshot). Image and video data are transmitted from all sensors to the FPGA via high-speed serial data lanes. The FPGA then organizes and routes this data to a standard desktop computer with 128 GB of RAM and a 4TB solid-state drive for permanent storage. As detailed below, the system enables full-frame video recording at seven frames per second and higher frame rates at lower per-frame pixel counts.

We applied a standard image stitching software to create all composite images post-capture unless stated otherwise. This customized software followed procedures currently available within the open-source Hugin code base~\cite{hugin}. In some experiments, it was beneficial to capture images of calibration targets to identify optimal stitching parameters before executing the imaging experiment. This pre-calibration step takes around two minutes for the first two imaging configurations and an hour for the tiled imaging configuration and then allows essentially instantaneous stitching of subsequent newly captured frames (as long as the sample is flat and at the same depth).

\subsection{Validation of optical resolution}

\begin{figure}[!t]
    \centering
    \includegraphics[scale = 0.355]{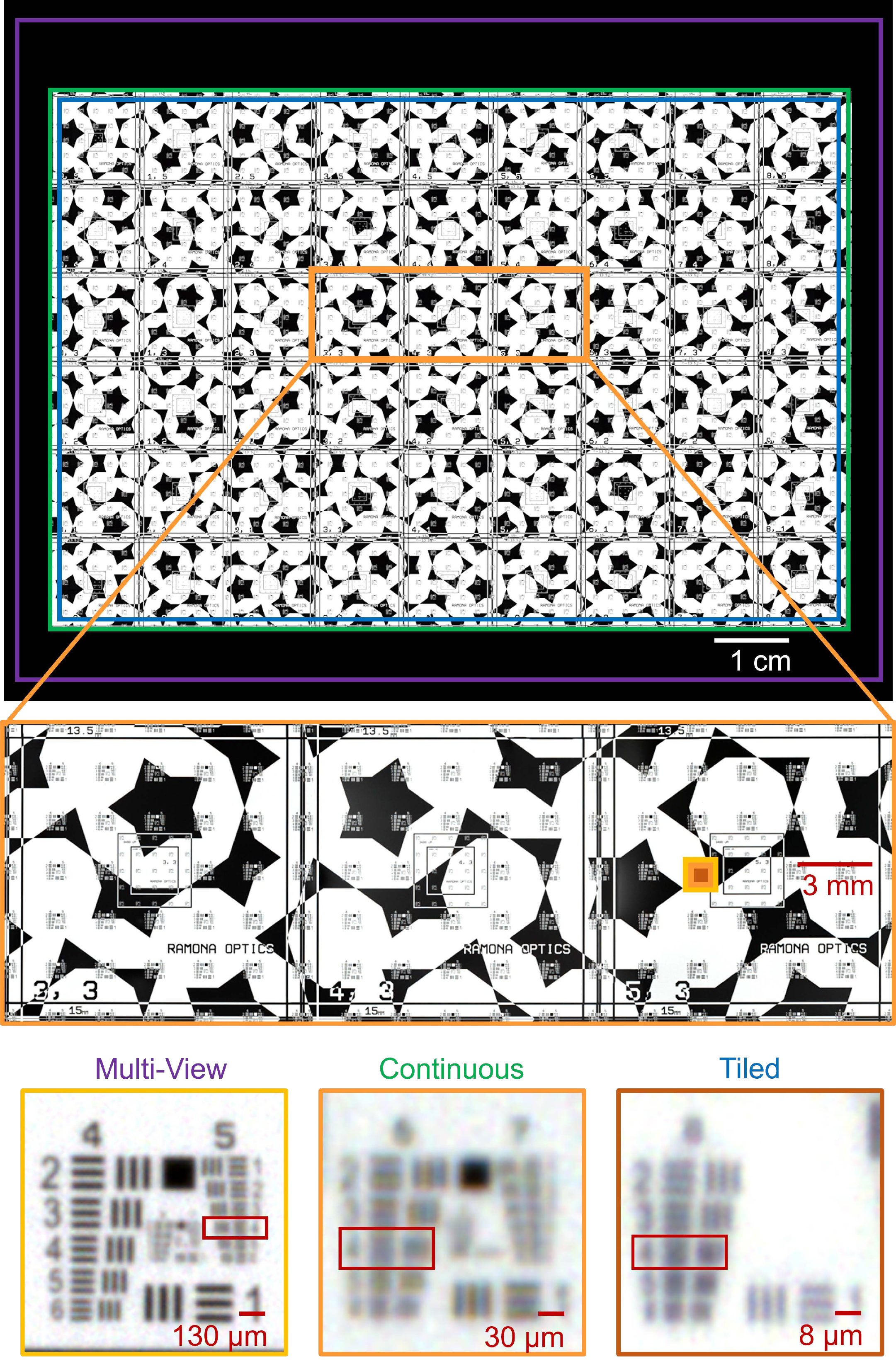}
    \caption{MCAM resolution and FOV assessment. Custom-designed resolution target imaged by three MCAM configurations (multi-view, continuous, tiled - labeled with purple, green, and blue, respectively). Gigapixels (GP) were reported for final stitched composites. Red boxes denote full-pitch resolutions of approximately 20, 10, and 2 $\mu$m.}
    \label{fig:threemodes_resolution}
\end{figure}

We summarize the results of a first set of experiments designed to assess MCAM system resolution in Figure~\ref{fig:threemodes_resolution}. We imaged a custom-designed resolution target with the printed area (green box in Figure~\ref{fig:threemodes_resolution}) spanning 83 x 123 mm² at three different working distances associated with each of the three configurations (multi-view, continuous, and tiled design). For multi-view imaging, a working distance of $WD_m=250$ mm yielded a magnification of $M_m=0.1$. For continuous imaging, a working distance of $WD_c=140$ mm yielded a magnification of $M_c=0.2$. Finally, for tiled imaging, we mounted a second array of 6 x 9 matching lenses atop the existing lens array but flipped it to produce a 4f imaging system for each micro-camera. This optical layout yielded an approximate working distance of $WD_t=5$ mm and magnification $M_t=1$, as seen in Table~\ref{tab:MCAM_specifications}.

\begin{table}[b]
\addtolength{\tabcolsep}{-4pt}
  \caption{\bf MCAM specifications}
  \label{tab:MCAM_specifications}
  \centering
  \begin{tabular}{lcccccc}
    \toprule
    Design & FOV (cm$^2$) & Res ($\mu$m) & M & WD (mm) & GP & \# snaps\\
    \midrule
     \textcolor{mypurple}{Multi-View} & 135 & 20 & 0.1 & 250 & 0.15 & 1 \\
     \textcolor{mygreen}{Continuous} & 100 & 10 & 0.2 & 140 & 0.48 & 1 \\
     \textcolor{myblue}{Tiled} & 96 & 2 & 1 & 5 & 9.8 & 25 \\
    \bottomrule
  \end{tabular}
\end{table}

The total array FOV varied slightly for each imaging configuration, which we denote with colored box outlines (purple, green, and blue) in Figure~\ref{fig:threemodes_resolution}. Additional key specifications for each design are highlighted in Table~\ref{tab:MCAM_specifications}. Maximum full-pitch resolution limits for each setup were 20 $\mu$m, 10 $\mu$m, and 2 $\mu$m for multi-view, continuous, and tiled imaging, respectively. The trend in achieved resolution follows the mathematical derivation in section~\ref{Multi-camera array microscope design}. We also demonstrate that the resolution changes very little across different camera positions and across different image regions of a single camera for the multi-view and continuous configurations, as shown in supplementary Figure S1. Tiled imaging (configuration 3), however, exhibits limited aberrations at the corner of each micro-camera FOV, as a working distance of the lenses utilized for this demonstration was $90-\infty$ mm, not the 5 mm used here. Future designs can utilize customized lens designs for tiled imaging, or crop such aberrated areas at the expense of requiring additional scanning. In the following subsections, we demonstrate MCAM imaging performance in each regime in a set of experiments.

\subsection{3D imaging with multi-view}
\begin{figure}[!t]
    \centering
    \includegraphics[scale = 0.3]{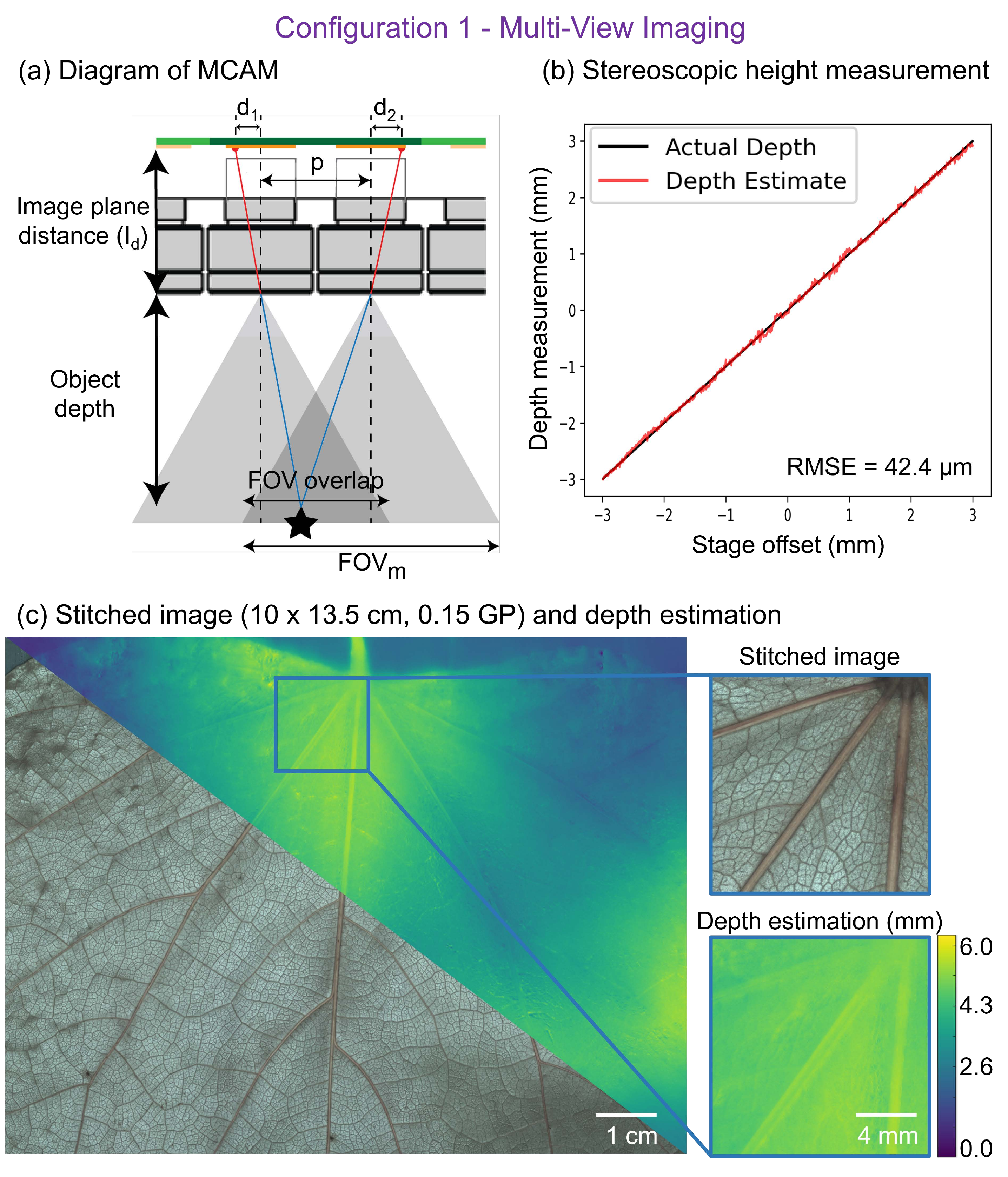}
    \caption{(a) Diagram of multi-view MCAM. FOV of each lens-sensor pair is $FOV_m=32$ mm. (b) Plot of stereoscopic height measurement accuracy (c) Example stitched composite with a jointly estimated 3D height map. Full gigapixel-scale results can be found at gigazoom.rc.duke.edu.}
    \label{fig:multiview_results}
\end{figure}

Stereo imaging is a standard technique in computer vision that can estimate object depth based on imaging from multiple views~\cite{zhang2021gigamvs}. Through a similar principle, the multi-view imaging configuration of the MCAM (with 50 \% or more overlap between the FOV of adjacent cameras) can be used to estimate the 3D height of specimens. The most direct explanation of stereoscopic-based depth estimation follows the principle of triangulation. As sketched in Figure~\ref{fig:multiview_results}(a), a point on the object plane will map to two image plane locations on two unique sensors in a multi-view setup configured with 50 \% overlap. Each imaged location will be at certain a lateral disparity, $d_1$ and $d_2$, with respect to the optical axes of camera 1 and 2. From these measurable lateral disparities and known optical setup parameters, such as inter-camera pitch $p$ and image plane distance $I_d$, object depth can be computed as,
\begin{equation}
Object\ depth = \frac{pI_d}{d_1 + d_2},
 \label{triangulation_equation}
\end{equation}
To verify this principle with our multi-view MCAM, we performed an experiment to measure the accuracy of this depth estimation by recording images of a standard USAF resolution target at a range of object distances. After calibrating the multi-view MCAM to exhibit slightly more than 50\% overlap in each dimension ($WD_m=250$ mm), we axially displaced the resolution target from -3 to 3 mm from the originally calibrated focal plane and captured images at 10 $\mu$m increments. From two acquired camera images per depth plane, we then used standard stereoscopic methods to 1) find common features across both images, 2) compute disparity distances $d_1$ and $d_2$ from corresponding optical axes, and 3) apply Eq.~\ref{triangulation_equation} to estimate object depth. As plotted on the y-axis in Figure~\ref{fig:multiview_results}(b), we observe accurate performance across the 6 mm range with an experimental RMSE of 42.4 $\mu$m with respect to ground-truth depth.


Photogrammetric 3D reconstruction algorithms can also be used to produce dense, pixel-wise 3D surface height maps that are co-registered with the stitched photometric images (Figure~\ref{fig:multiview_results}(c)). The problem of estimating 3D information from a collection of multi-view 2D images has been extensively studied in the computer vision community at macroscopic scales~\cite{ullman1979interpretation, schonberger2016structure} and more recently at smaller, mesoscopic (mm) scales~\cite{zhou2021mesoscopic}. Here, we adapted the algorithmic approach laid out in~\cite{zhou2021mesoscopic} to jointly reconstruct a stitched composite and 3D surface height map for all individual 54 MCAM sub-images. The central idea behind this computational procedure is to jointly match and register sample features viewed from multiple perspectives while converting parallax distortions to an estimated sample height. Unlike light-field microscopes, the multi-view MCAM can jointly provide a significantly enlarged image FOV with hundreds of resolved megapixels per snapshot, in addition to a 3D height map with an estimated axial sensitivity of 42 $\mu$m. Figure~\ref{fig:multiview_results}(c) demonstrates 3D surface height map formation for a leaf across a 135 cm$^2$ area, from which the 3D nature of the midrib is clearly highlighted.

\subsection{Gigapixel video over a continuous area}
\begin{figure}[!t]
    \centering
    \includegraphics[scale = 0.34]{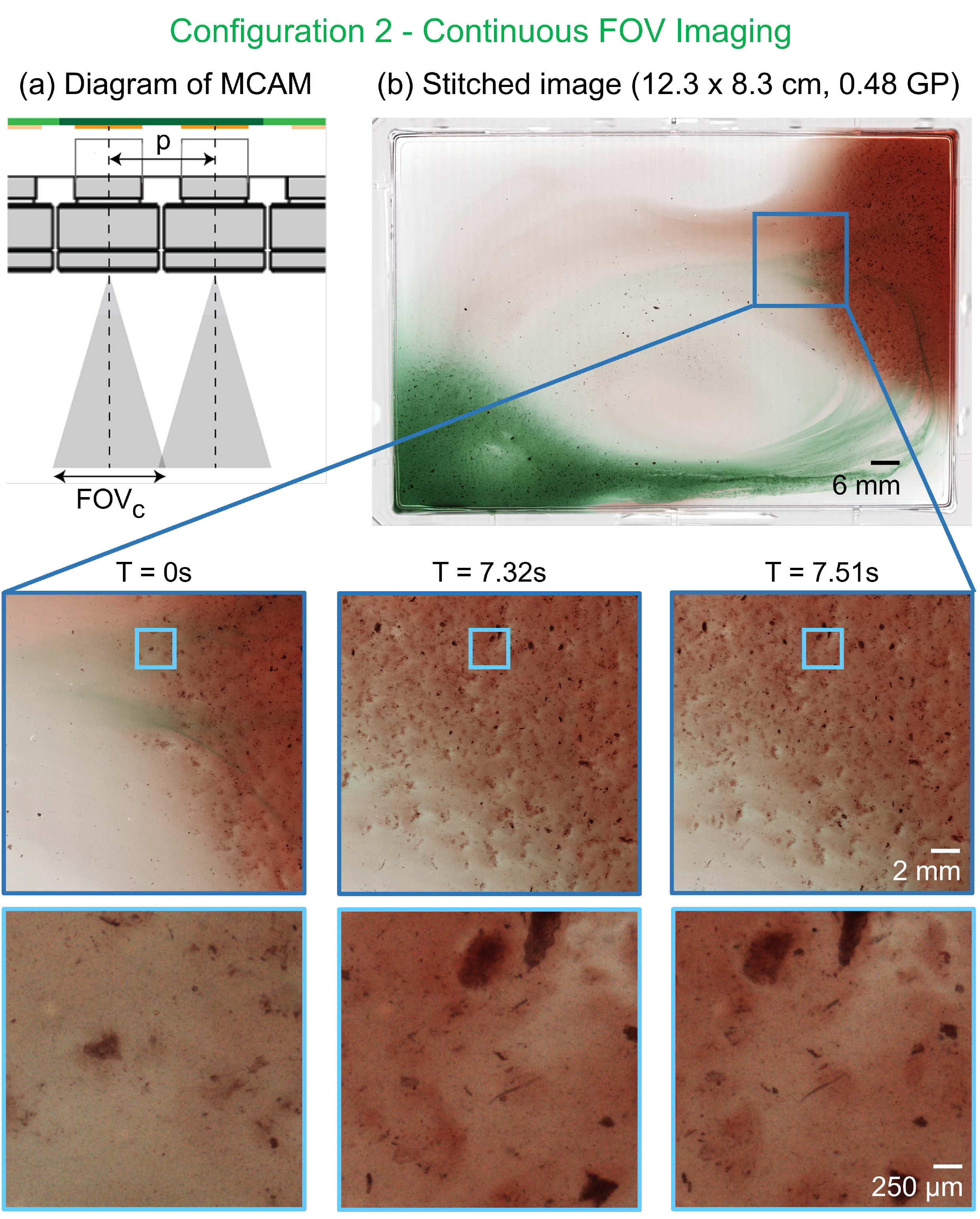}
    \caption{(a) Diagram of continuous FOV MCAM. FOV of each lens-sensor pair is $FOV_c=16$ mm. (b) Continuous FOV MCAM video of a dynamic fluid mixture. Full gigapixel-scale results can be found at gigazoom.rc.duke.edu. }
    \label{fig:continous_results}
\end{figure}

For the next demonstration of MCAM imaging, we arranged the multi-camera array for continuous area imaging at a working distance of $WD_c=140$ mm, which creates a small amount (approx. 5-10\%) of inter-camera overlap (Figure~\ref{fig:continous_results}(a)). As shown in Figure~\ref{fig:threemodes}, This configuration provides a higher magnification and spatial resolution for contiguous area imaging as compared to the multi-view arrangement.

A video of a viscous fluid mixture (areas dyed red and green for visualization) was recorded at a frame rate of 5 Hz for a duration of 21 seconds (115 frames were recorded within 128GB RAM). Two exemplary frames highlighting liquid movement are in Figure~\ref{fig:continous_results}(b) and an example recording is in Visualization 1, wherein the dynamics of small particulate matter is clearly observable. As detailed in the Discussion, current MCAM frame rates are limited by FPGA-to-workstation data transmission. Video data can be acquired at higher frame rates across the same contiguous area, by reducing per-frame data with on-chip pixel binning, wherein adjacent pixel values are combined pre-transmission for a reduced final frame pixel count. For example, with the CMOS sensors and electronics utilized in the current configuration, an approximate 28 frame rate is achieved with the use of 2X pixel binning, at the expense of a 2X lower pixel-limited resolution.

\subsection{9.8 gigapixel tiled imaging}
\begin{figure}[!t]
    \centering
    \includegraphics[scale = 0.308]{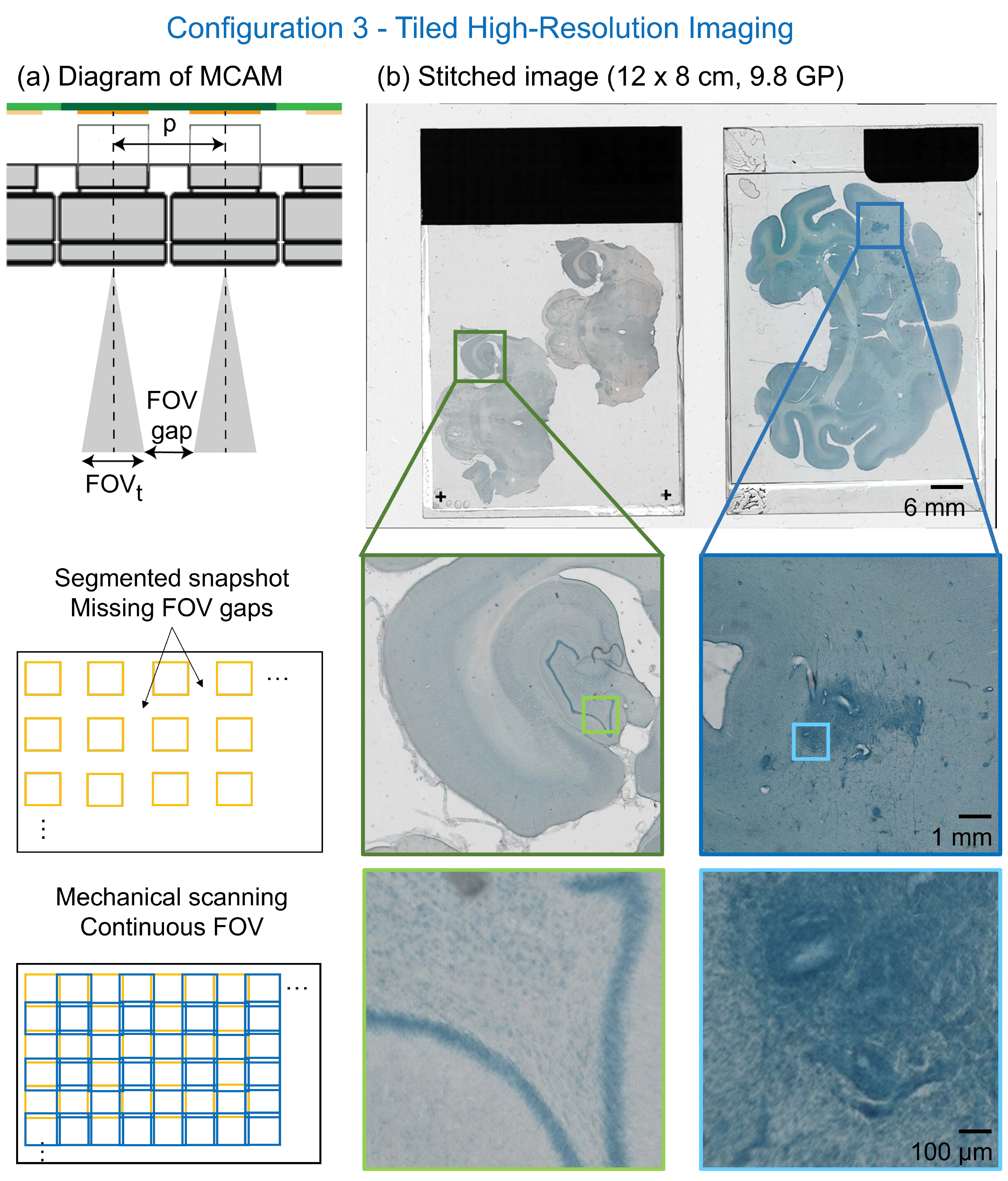}
    \caption{(a) Diagram of tiled high-resolution MCAM. FOV of each lens-sensor pair is $FOV_t=3.5$ mm. (b) Large-area simultaneous scan of multiple brain slices. Full gigapixel-scale results can be found at gigazoom.rc.duke.edu. }
    \label{fig:tiled_results}
\end{figure}

Finally, to demonstrate higher resolution MCAM imaging, we created a slightly modified optical arrangement and outfitted the specimen plane for limited translation scanning. For each of the 54 micro-cameras in the tiled imaging configuration, we arranged two of the same lenses used for multi-view and continuous imaging in a 4f configuration. This produced 54 unique 1x magnification images, each with an approximate 2 $\mu$m full-pitch resolution limit. The maximum lateral travel distance for MCAM tiled imaging is specified by $p$, the inter-camera spacing, which here is 13.5 mm Figure~\ref{fig:tiled_results}(a)). This particular tiled imaging setup requires 5 x 5 scans (25 snapshots) to cover the entire FOV, where each snapshot's FOV overlaps with adjacent snapshots by approximately 10\% for effective stitching. The resulting dataset from our 25 scans contains 1350 unique micro-camera images (13 megapixels per image).

Figure~\ref{fig:tiled_results}(b) shows a set of two macaque brain slices (75 $\mu$m thick) arranged on slides, which were simultaneously imaged using the tiled MCAM configuration to produce final composites with 9.8 GP in total. In this example, along with lateral scanning to fill in inter-camera FOV gaps, we also executed axial scanning (10 slices at 10 $\mu$m increment) to account for the uneven surface of the relatively thick tissue specimens. For each micro-camera at each lateral scan position, we selected the most in-focus image via a Laplacian-based contrast metric for final composite synthesis. The bottom-left images in Figure~\ref{fig:tiled_results}(b) show the CA2 region of the hippocampus, while the bottom-right images highlight an area exhibiting gliosis.

\begin{figure}
    \centering
    \includegraphics[scale = 0.32]{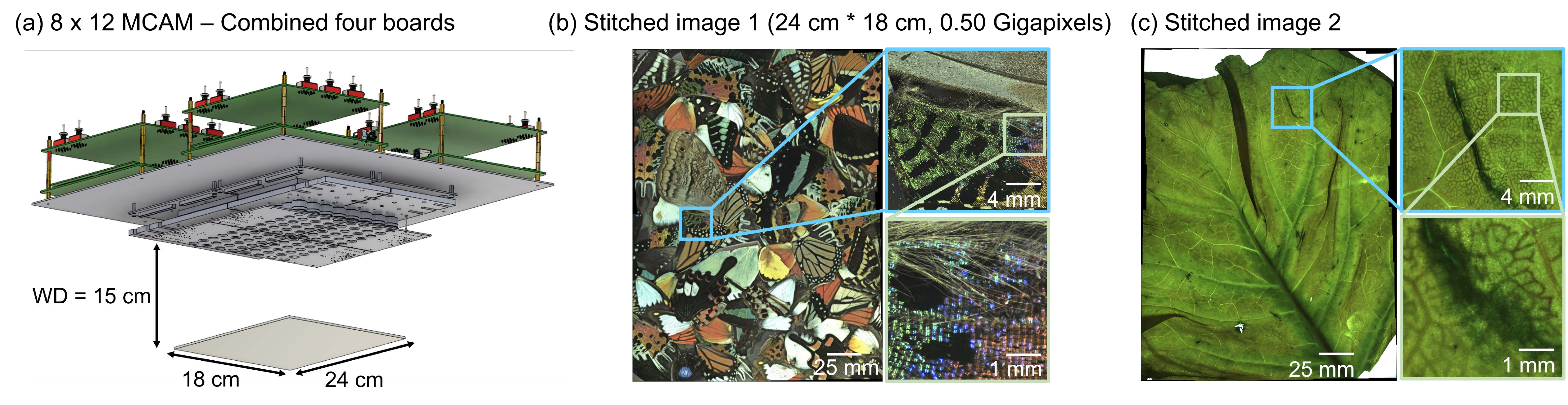}
    \caption{To demonstrate the straight-forward scalability of the MCAM system, four MCAMs (24 micro-cameras each, 10 megapixel sensors, $p=19$ mm pitch) were tiled together to image a 4X larger FOV. (a) Optical arrangement for acquiring example stitched composites with 0.96 gigapixels per snapshot of (b) large collection of butterfly wings and (c) collard green leaf.}
    \label{fig:extension_multipleboards}
\end{figure}

\section{Discussion and Conclusion}
The three configurations of the MCAM system, presented in this work are a natural starting point for exploring the utility of parallelized acquisition in microscopy. By providing gigapixel-scale single-snapshot throughput, the MCAM technology opens up new possibilities for wide-area high-resolution 3D imaging and video. There is a number of existing challenges and future directions for MCAM technology improvement that will likely lead to exciting follow-up work, which we summarize below.

One primary challenge faced by multi-aperture microscope designs is data management. In this work, we utilized a single FPGA to aggregate Mobile Industry Processor Interface (MIPI) data directly from all sensors, which led to several key benefits. First, the single FPGA allowed us to directly synchronize video capture across all 54 cameras within the array, leading to the ability to record gigapixel-scale video data without motion artifacts. Second, we were able to achieve a maximum data transmission rate of approximately 5 GB/sec, which corresponds to approximately 7 frames per second video (8 bits/pixel) with our 54-camera arrangement. It is possible to crop the sensors or reduce the camera count to increase the frame rate; for example, a 3072$\times$3072 square crop yields $\sim$10 frames per second. Transfer at such high data rates must account for limitations both in transmission links and in drive write speeds. In our prototype, we utilized a PCIe connection between the computer and the MCAM and stored data on pre-allocated RAM space before moving to a solid-state drive, which enabled approximately 249 seconds of video capture at full resolution. As solid-state storage volumes and write speeds continue to improve, we anticipate the ability to increase data transmission rates, and thus imaging frame rates, using the current array. We also anticipate that improved data transmission will open the door to novel MCAM designs that offer even larger full-frame sizes. Compression methods are likewise available to dramatically reduce video sizes in a lossless manner, for example~\cite{sullivan2012overview, mukherjee2013latest, chen2018overview}, and we see the FPGA as a key means to pre-process image data for additional future speed-up.

A second challenge is reaching a higher resolution. Following section~\ref{Multi-camera array microscope design}'s analysis, it is clear that resolution can be improved beyond our initial demonstrations. For tiled imaging, custom-designed lenses with superior magnification could match the resolution performance of standard high-NA objective lenses over the demonstrated 8 x 12 cm² FOV. For continuous and multi-view imaging, one primary driver of resolution is the inter-camera spacing $p$. Future MCAM designs can readily be created with $p < 13.5$mm for a proportional increase in resolution performance. Smaller pixels and/or alternative single-photon detector arrays~\cite{fossum2016quanta, morimoto2020megapixel, zappa2007principles} may also be adopted. Our demonstrated design likewise utilized 4:3 format CMOS sensors. It is alternatively possible to utilize perfectly square sensors for uniform overlap in x and y. Or, highly rectangular sensors can be utilized for multi-view imaging with an overlap in just one spatial dimension (i.e., to facilitate stereoscopic imaging with just two views per sample plane location), which may lead to improved resolution multi-view imaging. Finally, a non-rectangular (e.g., hexagonal) micro-camera packing geometry may also yield future performance gains.

It is also desirable to ensure that the macroscopic objects imaged here remain in sharp focus. We used standard calibration processes to first focus all 54 lenses within our prototype array for the three demonstrated configurations onto a common plane. However, afterward, care was required to ensure each macroscopic sample was placed and remained in-focus. As the MCAM transitions to higher resolution designs with shallower depths of field, several strategies may prove helpful. For example, a phase mask may be included for extended depth of field imaging~\cite{Dowski:95}. A per-lens autofocus capability would enable imaging at high focus across curved specimens, or even surfaces that variably change in height as a function of time.

In terms of imaging FOV, a particular MCAM design's coverage is simply proportional to the number of micro-cameras included within an array. If a larger FOV is required, then it is possible to directly join multiple MCAMs to form a larger aggregate array. An example of such an aggregate MCAM array to image over a significantly larger FOV is shown in Figure~\ref{fig:extension_multipleboards}(a). In this demonstration, we combined four identical, separately constructed 4 x 6 micro-camera arrays into an aggregate 8 x 12 array that contained 96 individual lens-sensor pairs (10 megapixels each, 0.96 GP per snapshot). As demonstrated in Figure~\ref{fig:extension_multipleboards}(b, c), such a configuration can provide high-resolution continuous image capture over a large, macroscopic area covering approximately 18 x 24 cm². Additional details about this design can be found in Ref.~\cite{Thomson:21}.

The MCAM can also be configured for a wide variety of alternative imaging configurations. We have recently demonstrated multi-aperture fluorescence imaging~\cite{Thomson:21}. Polarization~\cite{guo2020revealing}, phase contrast~\cite{yang2022multi} and aperture coding~\cite{cieslak2016coded} techniques can also be easily implemented to extract additional specimen information. Likewise, while it has been previously explored in tiled imaging configurations that do not image a continuous area~\cite{chan2019parallel}, variable-angle dark-field and bright-field illumination is an alternative direction for novel configurations of continuous and multi-view MCAM setups. As with all digital microscopes, software will continue to play an increasingly important role in the pre-processing and analysis of the MCAM’s terapixel-sized datasets. While this work performed all processing post-capture, future implementations may benefit from on-the-fly or hardware-accelerated image stitching and/or surface height estimation. For application-specific scenarios such as object tracking and feature detection, various methods may be used to significantly reduce acquired file sizes on-the-fly while still maintaining salient image information.

\section{Funding} Research reported in this publication was supported by the Office of Research Infrastructure Programs (ORIP), Office Of The Director, and the National Institute Of Environmental Health Sciences (NIEHS) of the National Institutes of Health under Award Number R44OD024879, by the National Cancer Institute (NCI) of the National Institutes of Health under Award Number R44CA250877, by the National Institute of Biomedical Imaging and Bioengineering (NIBIB) of the National Institutes of Health under Award Number R43EB030979, the National Science Foundation under Award Number 2036439, and the Duke Coulter Translational Partnership Award.

\section{Acknowledgments} The authors would like to thank Dr. Benjamin Judkewitz, Dr. Florian Engert and Dr. Michael Orger for helpful guidance during project development. 

\section{Disclosures} MH: Ramona Optics Inc. (F,I,P,S), KCZ: Ramona Optics Inc. (C), SS: Ramona Optics Inc. (C), CC: Ramona Optics Inc. (C), RA: Ramona Optics Inc. (I,E,P), CC: Ramona Optics Inc. (I,E,P), JD: Ramona Optics Inc. (I,E,P), GH: Ramona Optics Inc. (I,E,P), JP: Ramona Optics Inc. (I,E,P), PR: Ramona Optics Inc. (I,E,P), VS: Ramona Optics Inc. (I,E,P), RH: Ramona Optics Inc. (F,I,P,S).

\section{Data availability} Data underlying the results may be obtained from the authors upon reasonable request.

\section{Supplemental document}
See Supplement 1 for supporting content.

\bibliographystyle{unsrt}  
\bibliography{references}

\begin{thebibliography}{10}

\bibitem{zheng2014fourier}
Guoan Zheng, Xiaoze Ou, Roarke Horstmeyer, Jaebum Chung, and Changhuei Yang.
\newblock Fourier ptychographic microscopy: A gigapixel superscope for
  biomedicine.
\newblock {\em Optics and Photonics News}, 25(4):26--33, 2014.

\bibitem{lohmann1996space}
Adolf~W Lohmann, Rainer~G Dorsch, David Mendlovic, Zeev Zalevsky, and Carlos
  Ferreira.
\newblock Space--bandwidth product of optical signals and systems.
\newblock {\em JOSA A}, 13(3):470--473, 1996.

\bibitem{park2021review}
Jongchan Park, David~J Brady, Guoan Zheng, Lei Tian, and Liang Gao.
\newblock Review of bio-optical imaging systems with a high space-bandwidth
  product.
\newblock {\em Advanced Photonics}, 3(4):044001, 2021.

\bibitem{kim2017pan}
Dal~Hyung Kim, Jungsoo Kim, Jo{\~a}o~C Marques, Abhinav Grama, David~GC
  Hildebrand, Wenchao Gu, Jennifer~M Li, and Drew~N Robson.
\newblock Pan-neuronal calcium imaging with cellular resolution in freely
  swimming zebrafish.
\newblock {\em Nature methods}, 14(11):1107--1114, 2017.

\bibitem{nguyen2016whole}
Jeffrey~P Nguyen, Frederick~B Shipley, Ashley~N Linder, George~S Plummer, Mochi
  Liu, Sagar~U Setru, Joshua~W Shaevitz, and Andrew~M Leifer.
\newblock Whole-brain calcium imaging with cellular resolution in freely
  behaving caenorhabditis elegans.
\newblock {\em Proceedings of the National Academy of Sciences},
  113(8):E1074--E1081, 2016.

\bibitem{johnson2020probabilistic}
Robert~Evan Johnson, Scott Linderman, Thomas Panier, Caroline~Lei Wee, Erin
  Song, Kristian~Joseph Herrera, Andrew Miller, and Florian Engert.
\newblock Probabilistic models of larval zebrafish behavior reveal structure on
  many scales.
\newblock {\em Current Biology}, 30(1):70--82, 2020.

\bibitem{krishnamurthy2020scale}
Deepak Krishnamurthy, Hongquan Li, Fran{\c{c}}ois~Benoit du~Rey, Pierre
  Cambournac, Adam~G Larson, Ethan Li, and Manu Prakash.
\newblock Scale-free vertical tracking microscopy.
\newblock {\em Nature Methods}, 17(10):1040--1051, 2020.

\bibitem{acciani2006application}
Giuseppe Acciani, Gioacchino Brunetti, and Girolamo Fornarelli.
\newblock Application of neural networks in optical inspection and
  classification of solder joints in surface mount technology.
\newblock {\em IEEE Transactions on industrial informatics}, 2(3):200--209,
  2006.

\bibitem{baek2004inspection}
Seung-Il Baek, Woo-Seob Kim, Tak-Mo Koo, Il~Choi, and Kil-Houm Park.
\newblock Inspection of defect on lcd panel using polynomial approximation.
\newblock In {\em 2004 IEEE Region 10 Conference TENCON 2004.}, pages 235--238.
  IEEE, 2004.

\bibitem{wu2014wafer}
Ming-Ju Wu, Jyh-Shing~R Jang, and Jui-Long Chen.
\newblock Wafer map failure pattern recognition and similarity ranking for
  large-scale data sets.
\newblock {\em IEEE Transactions on Semiconductor Manufacturing}, 28(1):1--12,
  2014.

\bibitem{castiaux2019review}
Andre~D Castiaux, Dana~M Spence, and R~Scott Martin.
\newblock Review of 3d cell culture with analysis in microfluidic systems.
\newblock {\em Analytical Methods}, 11(33):4220--4232, 2019.

\bibitem{chen2020solar}
Haiyong Chen, Yue Pang, Qidi Hu, and Kun Liu.
\newblock Solar cell surface defect inspection based on multispectral
  convolutional neural network.
\newblock {\em Journal of Intelligent Manufacturing}, 31(2):453--468, 2020.

\bibitem{yun2020automated}
Jong~Pil Yun, Woosang~Crino Shin, Gyogwon Koo, Min~Su Kim, Chungki Lee, and
  Sang~Jun Lee.
\newblock Automated defect inspection system for metal surfaces based on deep
  learning and data augmentation.
\newblock {\em Journal of Manufacturing Systems}, 55:317--324, 2020.

\bibitem{Boutros2015}
Michael Boutros, Florian Heigwer, and Christina Laufer.
\newblock Microscopy-based high-content screening.
\newblock {\em Cell}, 163(6):1314--1325, 2015.

\bibitem{lohmann1989scaling}
Adolf~W Lohmann.
\newblock Scaling laws for lens systems.
\newblock {\em Applied optics}, 28(23):4996--4998, 1989.

\bibitem{Brady:18parallel}
David~J. Brady, Wubin Pang, Han Li, Zhan Ma, Yue Tao, and Xun Cao.
\newblock Parallel cameras.
\newblock {\em Optica}, 5(2):127--137, Feb 2018.

\bibitem{schniete2018fast}
Jan Schniete, Aimee Franssen, John Dempster, Trevor~J Bushell, William~Bradshaw
  Amos, and Gail McConnell.
\newblock Fast optical sectioning for widefield fluorescence mesoscopy with the
  mesolens based on hilo microscopy.
\newblock {\em Scientific reports}, 8(1):1--10, 2018.

\bibitem{canon_global}
Canon develops aps-h-size cmos sensor with approximately 250 megapixels, the
  world's highest pixel count for its size.

\bibitem{farahani2015whole}
Navid Farahani, Anil~V Parwani, Liron Pantanowitz, et~al.
\newblock Whole slide imaging in pathology: advantages, limitations, and
  emerging perspectives.
\newblock {\em Pathol Lab Med Int}, 7(23-33):4321, 2015.

\bibitem{chan201996}
Antony~CS Chan, Jinho Kim, An~Pan, Han Xu, Dana Nojima, Christopher Hale,
  Songli Wang, and Changhuei Yang.
\newblock 96 eyes: Parallel fourier ptychographic microscopy for
  high-throughput screening.
\newblock {\em bioRxiv}, page 547265, 2019.

\bibitem{maioli2016time}
Vincent Maioli, George Chennell, Hugh Sparks, Tobia Lana, Sunil Kumar, David
  Carling, Alessandro Sardini, and Chris Dunsby.
\newblock Time-lapse 3-d measurements of a glucose biosensor in multicellular
  spheroids by light sheet fluorescence microscopy in commercial 96-well
  plates.
\newblock {\em Scientific reports}, 6(1):1--13, 2016.

\bibitem{gustafsson2000surpassing}
Mats~GL Gustafsson.
\newblock Surpassing the lateral resolution limit by a factor of two using
  structured illumination microscopy.
\newblock {\em Journal of microscopy}, 198(2):82--87, 2000.

\bibitem{zheng2013wide}
Guoan Zheng, Roarke Horstmeyer, and Changhuei Yang.
\newblock Wide-field, high-resolution fourier ptychographic microscopy.
\newblock {\em Nature photonics}, 7(9):739--745, 2013.

\bibitem{Konda:20FPReview}
Pavan~Chandra Konda, Lars Loetgering, Kevin~C. Zhou, Shiqi Xu, Andrew~R.
  Harvey, and Roarke Horstmeyer.
\newblock Fourier ptychography: current applications and future promises.
\newblock {\em Opt. Express}, 28(7):9603--9630, Mar 2020.

\bibitem{mudry2012structured}
Emeric Mudry, Kamal Belkebir, J~Girard, Julien Savatier, Emmeran Le~Moal,
  C~Nicoletti, Marc Allain, and Anne Sentenac.
\newblock Structured illumination microscopy using unknown speckle patterns.
\newblock {\em Nature Photonics}, 6(5):312--315, 2012.

\bibitem{rust2006sub}
Michael~J Rust, Mark Bates, and Xiaowei Zhuang.
\newblock Sub-diffraction-limit imaging by stochastic optical reconstruction
  microscopy (storm).
\newblock {\em Nature methods}, 3(10):793--796, 2006.

\bibitem{manley2008high}
Suliana Manley, Jennifer~M Gillette, George~H Patterson, Hari Shroff, Harald~F
  Hess, Eric Betzig, and Jennifer Lippincott-Schwartz.
\newblock High-density mapping of single-molecule trajectories with
  photoactivated localization microscopy.
\newblock {\em Nature methods}, 5(2):155--157, 2008.

\bibitem{Orth:12}
Antony Orth and Kenneth Crozier.
\newblock Microscopy with microlens arrays: high throughput, high resolution
  and light-field imaging.
\newblock {\em Opt. Express}, 20(12):13522--13531, Jun 2012.

\bibitem{Pang:12}
Shuo Pang, Chao Han, Mihoko Kato, Paul~W. Sternberg, and Changhuei Yang.
\newblock Wide and scalable field-of-view talbot-grid-based fluorescence
  microscopy.
\newblock {\em Opt. Lett.}, 37(23):5018--5020, Dec 2012.

\bibitem{ashraf2021random}
Mishal Ashraf, Sharika Mohanan, Byu~Ri Sim, Anthony Tam, Kiamehr Rahemipour,
  Denis Brousseau, Simon Thibault, Alexander~D Corbett, and Gil Bub.
\newblock Random access parallel microscopy.
\newblock {\em Elife}, 10:e56426, 2021.

\bibitem{brady2012multiscale}
David~J Brady, Michael~E Gehm, Ronald~A Stack, Daniel~L Marks, David~S Kittle,
  Dathon~R Golish, EM~Vera, and Steven~D Feller.
\newblock Multiscale gigapixel photography.
\newblock {\em Nature}, 486(7403):386--389, 2012.

\bibitem{tanida2001thin}
Jun Tanida, Tomoya Kumagai, Kenji Yamada, Shigehiro Miyatake, Kouichi Ishida,
  Takashi Morimoto, Noriyuki Kondou, Daisuke Miyazaki, and Yoshiki Ichioka.
\newblock Thin observation module by bound optics (tombo): concept and
  experimental verification.
\newblock {\em Applied optics}, 40(11):1806--1813, 2001.

\bibitem{wilburn2005high}
Bennett Wilburn, Neel Joshi, Vaibhav Vaish, Eino-Ville Talvala, Emilio Antunez,
  Adam Barth, Andrew Adams, Mark Horowitz, and Marc Levoy.
\newblock High performance imaging using large camera arrays.
\newblock In {\em ACM SIGGRAPH 2005 Papers}, pages 765--776. 2005.

\bibitem{Yuan:21}
Xiaoyun Yuan, Mengqi Ji, Jiamin Wu, David~J. Brady, Qionghai Dai, and Lu~Fang.
\newblock A modular hierarchical array camera.
\newblock {\em Light Sci. App.}, 10(37):1--9, 2021.

\bibitem{Brady:09}
David~J. Brady and Nathan Hagen.
\newblock Multiscale lens design.
\newblock {\em Opt. Express}, 17(13):10659--10674, Jun 2009.

\bibitem{Tremblay:12}
Eric~J. Tremblay, Daniel~L. Marks, David~J. Brady, and Joseph~E. Ford.
\newblock Design and scaling of monocentric multiscale imagers.
\newblock {\em Appl. Opt.}, 51(20):4691--4702, Jul 2012.

\bibitem{marks2014characterization}
Daniel~L Marks, PR~Llull, Z~Phillips, JG~Anderson, SD~Feller, EM~Vera, HS~Son,
  S-H Youn, Jungsang Kim, ME~Gehm, et~al.
\newblock Characterization of the aware 10 two-gigapixel wide-field-of-view
  visible imager.
\newblock {\em Applied optics}, 53(13):C54--C63, 2014.

\bibitem{Schuster:19}
Glenn~M. Schuster, Donald~G. Dansereau, Gordon Wetzstein, and Joseph~E. Ford.
\newblock Panoramic single-aperture multi-sensor light field camera.
\newblock {\em Opt. Express}, 27(26):37257--37273, Dec 2019.

\bibitem{fan2019video}
Jingtao Fan, Jinli Suo, Jiamin Wu, Hao Xie, Yibing Shen, Feng Chen, Guijin
  Wang, Liangcai Cao, Guofan Jin, Quansheng He, et~al.
\newblock Video-rate imaging of biological dynamics at centimetre scale and
  micrometre resolution.
\newblock {\em Nature Photonics}, 13(11):809--816, 2019.

\bibitem{konda2018parallelized}
Pavan~Chandra Konda, Jonathan~M Taylor, and Andrew~R Harvey.
\newblock Parallelized aperture synthesis using multi-aperture fourier
  ptychographic microscopy.
\newblock {\em arXiv preprint arXiv:1806.02317}, 2018.

\bibitem{kim20205}
HyunChul Kim, Jongeun Park, Insung Joe, Doowon Kwon, Joo~Hyoung Kim, Dongsuk
  Cho, Taehun Lee, Changkyu Lee, Haeyong Park, Soojin Hong, et~al.
\newblock A 1/2.65 in 44mpixel cmos image sensor with 0.7 $\mu$m pixels
  fabricated in advanced full-depth deep-trench isolation technology.
\newblock In {\em 2020 IEEE International Solid-State Circuits
  Conference-(ISSCC)}, pages 104--106. IEEE, 2020.

\bibitem{fossum2016quanta}
Eric~R Fossum, Jiaju Ma, Saleh Masoodian, Leo Anzagira, and Rachel Zizza.
\newblock The quanta image sensor: Every photon counts.
\newblock {\em Sensors}, 16(8):1260, 2016.

\bibitem{marks2011close}
Daniel~L Marks and David~J Brady.
\newblock Close-up imaging using microcamera arrays for focal plane synthesis.
\newblock {\em Optical Engineering}, 50(3):033205, 2011.

\bibitem{levoy2006light}
Marc Levoy, Ren Ng, Andrew Adams, Matthew Footer, and Mark Horowitz.
\newblock Light field microscopy.
\newblock In {\em ACM SIGGRAPH 2006 Papers}, SIGGRAPH '06, page 924–934, New
  York, NY, USA, 2006. Association for Computing Machinery.

\bibitem{broxton2013wave}
Michael Broxton, Logan Grosenick, Samuel Yang, Noy Cohen, Aaron Andalman, Karl
  Deisseroth, and Marc Levoy.
\newblock Wave optics theory and 3-d deconvolution for the light field
  microscope.
\newblock {\em Optics express}, 21(21):25418--25439, 2013.

\bibitem{prevedel2014simultaneous}
Robert Prevedel, Young-Gyu Yoon, Maximilian Hoffmann, Nikita Pak, Gordon
  Wetzstein, Saul Kato, Tina Schr{\"o}del, Ramesh Raskar, Manuel Zimmer,
  Edward~S Boyden, et~al.
\newblock Simultaneous whole-animal 3d imaging of neuronal activity using
  light-field microscopy.
\newblock {\em Nature methods}, 11(7):727--730, 2014.

\bibitem{lin2015camera}
Xing Lin, Jiamin Wu, Guoan Zheng, and Qionghai Dai.
\newblock Camera array based light field microscopy.
\newblock {\em Biomedical optics express}, 6(9):3179--3189, 2015.

\bibitem{pei2019occluded}
Zhao Pei, Yawen Li, Miao Ma, Jun Li, Chengcai Leng, Xiaoqiang Zhang, and
  Yanning Zhang.
\newblock Occluded-object 3d reconstruction using camera array synthetic
  aperture imaging.
\newblock {\em Sensors}, 19(3):607, 2019.

\bibitem{hugin}
Pablo d’Angelo et~al.
\newblock Hugin-panorama photo stitcher.
\newblock {\em Sourceforge. net. http://hugin. sourceforge. net/.(accessed May
  15, 2021)}, 2021.

\bibitem{zhang2021gigamvs}
Jianing Zhang, Jinzhi Zhang, Shi Mao, Mengqi Ji, Guangyu Wang, Zequn Chen, Tian
  Zhang, Xiaoyun Yuan, Qionghai Dai, and Lu~Fang.
\newblock Gigamvs: A benchmark for ultra-large-scale gigapixel-level 3d
  reconstruction.
\newblock {\em IEEE Transactions on Pattern Analysis \& Machine Intelligence},
  (01):1--1, 2021.

\bibitem{ullman1979interpretation}
Shimon Ullman.
\newblock The interpretation of structure from motion.
\newblock {\em Proceedings of the Royal Society of London. Series B. Biological
  Sciences}, 203(1153):405--426, 1979.

\bibitem{schonberger2016structure}
Johannes~L Schonberger and Jan-Michael Frahm.
\newblock Structure-from-motion revisited.
\newblock In {\em Proceedings of the IEEE conference on computer vision and
  pattern recognition}, pages 4104--4113, 2016.

\bibitem{zhou2021mesoscopic}
Kevin~C Zhou, Colin Cooke, Jaehee Park, Ruobing Qian, Roarke Horstmeyer,
  Joseph~A Izatt, and Sina Farsiu.
\newblock Mesoscopic photogrammetry with an unstabilized phone camera.
\newblock In {\em Proceedings of the IEEE/CVF Conference on Computer Vision and
  Pattern Recognition}, pages 7535--7545, 2021.

\bibitem{sullivan2012overview}
Gary~J Sullivan, Jens-Rainer Ohm, Woo-Jin Han, and Thomas Wiegand.
\newblock Overview of the high efficiency video coding (hevc) standard.
\newblock {\em IEEE Transactions on circuits and systems for video technology},
  22(12):1649--1668, 2012.

\bibitem{mukherjee2013latest}
Debargha Mukherjee, Jim Bankoski, Adrian Grange, Jingning Han, John Koleszar,
  Paul Wilkins, Yaowu Xu, and Ronald Bultje.
\newblock The latest open-source video codec vp9-an overview and preliminary
  results.
\newblock In {\em 2013 Picture Coding Symposium (PCS)}, pages 390--393. IEEE,
  2013.

\bibitem{chen2018overview}
Yue Chen, Debargha Murherjee, Jingning Han, Adrian Grange, Yaowu Xu, Zoe Liu,
  Sarah Parker, Cheng Chen, Hui Su, Urvang Joshi, et~al.
\newblock An overview of core coding tools in the av1 video codec.
\newblock In {\em 2018 Picture Coding Symposium (PCS)}, pages 41--45. IEEE,
  2018.

\bibitem{morimoto2020megapixel}
Kazuhiro Morimoto, Andrei Ardelean, Ming-Lo Wu, Arin~Can Ulku, Ivan~Michel
  Antolovic, Claudio Bruschini, and Edoardo Charbon.
\newblock Megapixel time-gated spad image sensor for 2d and 3d imaging
  applications.
\newblock {\em Optica}, 7(4):346--354, 2020.

\bibitem{zappa2007principles}
Franco Zappa, Simone Tisa, Alberto Tosi, and Sergio Cova.
\newblock Principles and features of single-photon avalanche diode arrays.
\newblock {\em Sensors and Actuators A: Physical}, 140(1):103--112, 2007.

\bibitem{Dowski:95}
Edward~R. Dowski and W.~Thomas Cathey.
\newblock Extended depth of field through wave-front coding.
\newblock {\em Appl. Opt.}, 34(11):1859--1866, Apr 1995.

\bibitem{Thomson:21}
Eric~E Thomson, Mark Harfouche, Pavan~C Konda, Catherine Seitz, Kanghyun Kim,
  Colin Cooke, Shiqi Xu, Robin Blazing, Yang~Chen ad~Whitney~Jacobs, Sunanda
  Sharma, Timothy~W Dunn, Jaehee Park, Roarke Horstmeyer, and Eva~A Naumann.
\newblock Gigapixel behavioral and neural activity imaging with a novel
  multi-camera array microscope.
\newblock {\em bioRxiv}, 2021.

\bibitem{guo2020revealing}
Syuan-Ming Guo, Li-Hao Yeh, Jenny Folkesson, Ivan~E Ivanov, Anitha~P Krishnan,
  Matthew~G Keefe, Ezzat Hashemi, David Shin, Bryant~B Chhun, Nathan~H Cho,
  et~al.
\newblock Revealing architectural order with quantitative label-free imaging
  and deep learning.
\newblock {\em elife}, 9:e55502, 2020.

\bibitem{yang2022multi}
Xi~Yang, Mark Harfouche, Kevin~C Zhou, Lucas Kreiss, Shiqi Xu, Kanghyun Kim,
  and Roarke Horstmeyer.
\newblock Multi-modal imaging using a cascaded microscope design.
\newblock {\em arXiv preprint arXiv:2208.08875}, 2022.

\bibitem{cieslak2016coded}
Micha{\l}~J Cie{\'s}lak, Kelum~AA Gamage, and Robert Glover.
\newblock Coded-aperture imaging systems: Past, present and future
  development--a review.
\newblock {\em Radiation Measurements}, 92:59--71, 2016.

\bibitem{chan2019parallel}
Antony~CS Chan, Jinho Kim, An~Pan, Han Xu, Dana Nojima, Christopher Hale,
  Songli Wang, and Changhuei Yang.
\newblock Parallel fourier ptychographic microscopy for high-throughput
  screening with 96 cameras (96 eyes).
\newblock {\em Scientific reports}, 9(1):1--12, 2019.

\end{thebibliography}

\appendix
\section{Supplementary Figure S1 Resolution analysis}
\begin{figure}[htbp]
\centering
\fbox{\includegraphics[width=.55\linewidth]{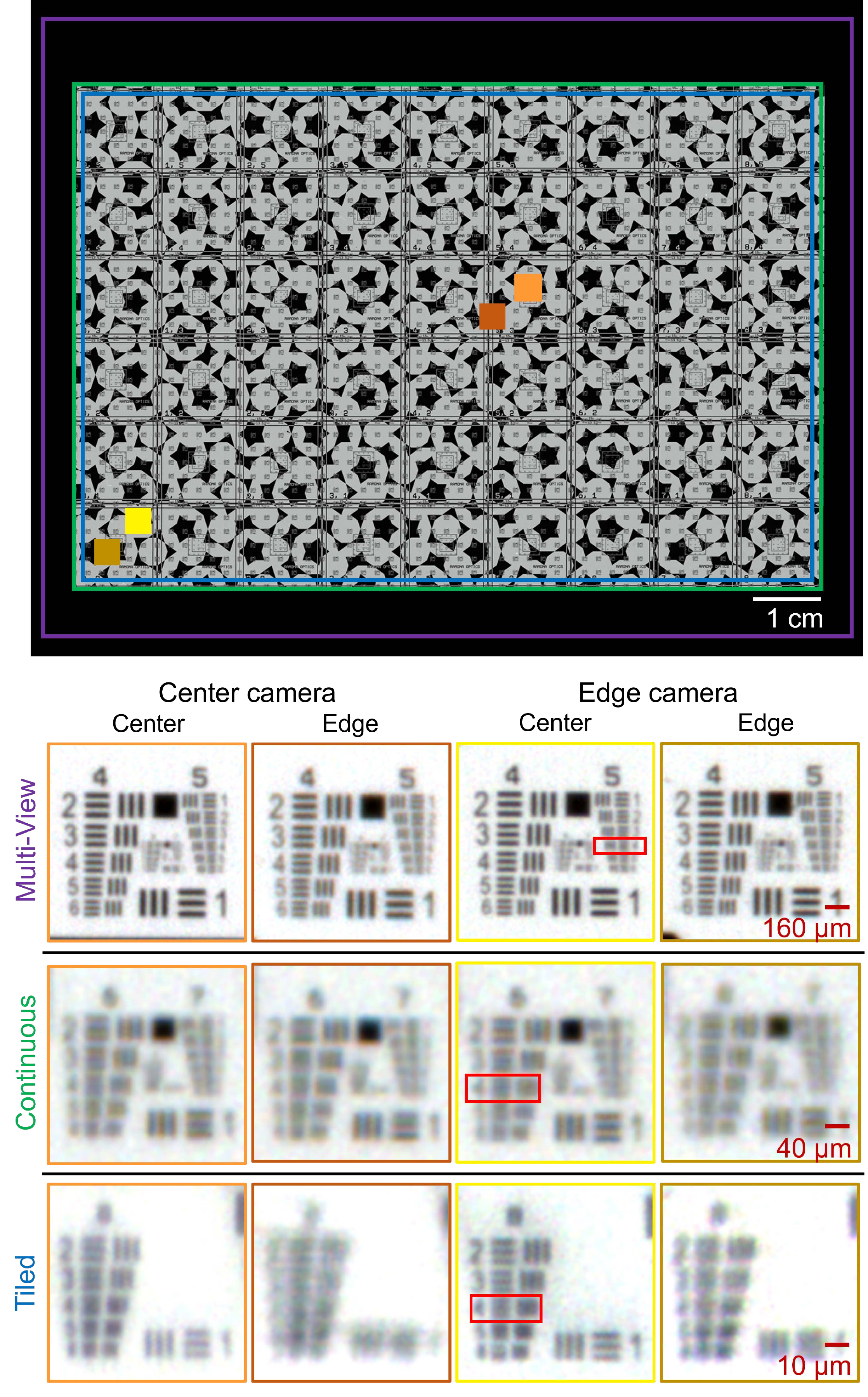}}
\caption{Experimental results of MCAM resolution assessment across camera and FOV. Zoom-ins at four marked locations in Figure 3 for each MCAM configuration. Red boxes denote full-pitch resolutions of approximately 20, 10, and 2 $\mu$m.}
\label{fig:false-color}
\end{figure}

\end{document}